\documentclass[sigconf]{acmart}
\usepackage{multirow}
\usepackage{tabularx}
\usepackage{fancyhdr}
\usepackage{booktabs} 
\usepackage[absolute,overlay]{textpos}
\AtBeginDocument{%
  }

\begin{document}

\begin{textblock*}{\paperwidth}(1.5cm, 1.2cm)
\includegraphics[height=0.7cm]{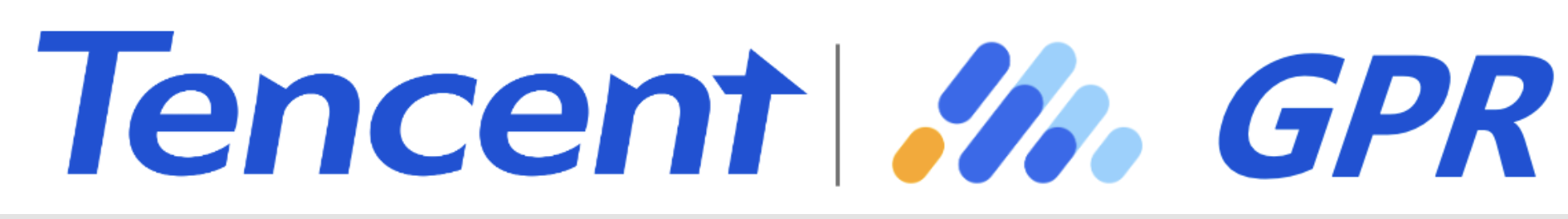} 
\end{textblock*}

\title{Recurrent Preference Memory for Efficient Long-Sequence Generative Recommendation}

\author{{\Large Yixiao Chen $^{1}$, Yuan Wang $^{2}$, Yue Liu $^{2}$, Qiyao Wang $^{2}$, Ke Cheng $^{2}$, Xin Xu $^{2}$, Juntong Yan $^{2}$, Shuojin Yang $^{1}$, Meng-Hao Guo $^{1}$, Jun Zhang $^{2}$, Huan Yu $^{2}$, Jie Jiang $^{2}$}}

\affiliation{
  \institution{ $^1$ Tsinghua University, China; $^2$ Tencent Inc., China}
  \country{}
}

\renewcommand{\shortauthors}{Chen et al.}

\settopmatter{printacmref=false} 
\renewcommand\footnotetextcopyrightpermission[1]{}

\begin{abstract}
Generative recommendation (GenRec) models typically model user behavior via full attention, but scaling to lifelong sequences is hindered by prohibitive computational costs and noise accumulation from stochastic interactions. To address these challenges, we introduce \textbf{Rec2PM}, a framework that compresses long user interaction histories into compact \textit{Preference Memory} tokens. Unlike traditional recurrent methods that suffer from serial training, Rec2PM employs a novel \textit{self-referential teacher-forcing} strategy: it leverages a global view of the history to generate ``reference memories,'' which serve as supervision targets for parallelized recurrent updates. This allows for fully parallel training while maintaining the capability for iterative updates during inference. Additionally, by representing memory as token embeddings rather than extensive KV caches, Rec2PM achieves extreme storage efficiency. Experiments on large-scale benchmarks show that Rec2PM significantly reduces inference latency and memory footprint while achieving superior accuracy compared to full-sequence models. Analysis reveals that the Preference Memory functions as a denoising \textit{Information Bottleneck}, effectively filtering interaction noise to capture robust long-term interests.
\end{abstract}





\received{20 February 2007}
\received[revised]{12 March 2009}
\received[accepted]{5 June 2009}

\maketitle

\section{Introduction}

Generative Recommendation (GenRec) has emerged as a promising paradigm, leveraging the Transformer architecture to model user behavior as a sequential generation task~\cite{Zhang2025GPR, zhai2024actions}. By capturing complex dependencies within historical interaction sequences, GenRec demonstrates strong capability in predicting the next item a user is likely to interact with. 

However, scaling GenRec to \emph{lifelong} user histories remains fundamentally challenging: interaction sequences can easily span thousands or even tens of thousands of tokens.

Long histories create two coupled issues. \textbf{First}, the quadratic cost of standard self-attention ($O(L^2)$) makes full life-cycle modeling prohibitively expensive in industrial serving, forcing systems to truncate to a short recent window and lose long-term preference signals. \textbf{Second}, user behaviors are stochastic and noisy (e.g., accidental clicks). Even if full-context computation were feasible, directly attending to the entire raw history may amplify such noise, distracting the model from the underlying preference structure and consequently weakening its generalization ability.

\begin{figure}[t]
    \centering
    \includegraphics[width=1.\linewidth]{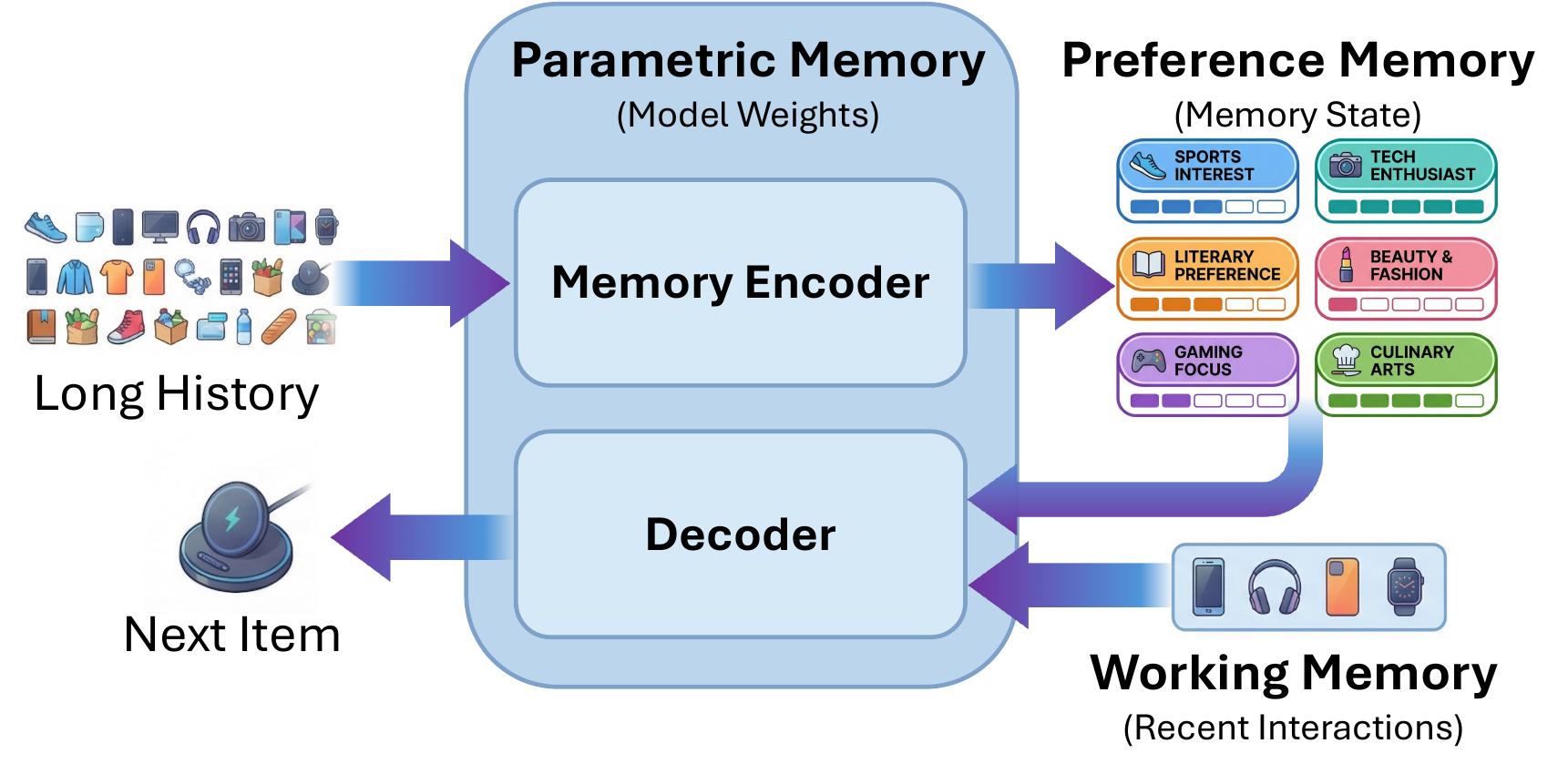}
    \caption{A generative recommendation system constructed based on a Tripartite Memory Mechanism.}
    \label{fig:CLS}
\end{figure}

\begin{figure*}[t]
    \centering
    \includegraphics[width=0.8\linewidth]{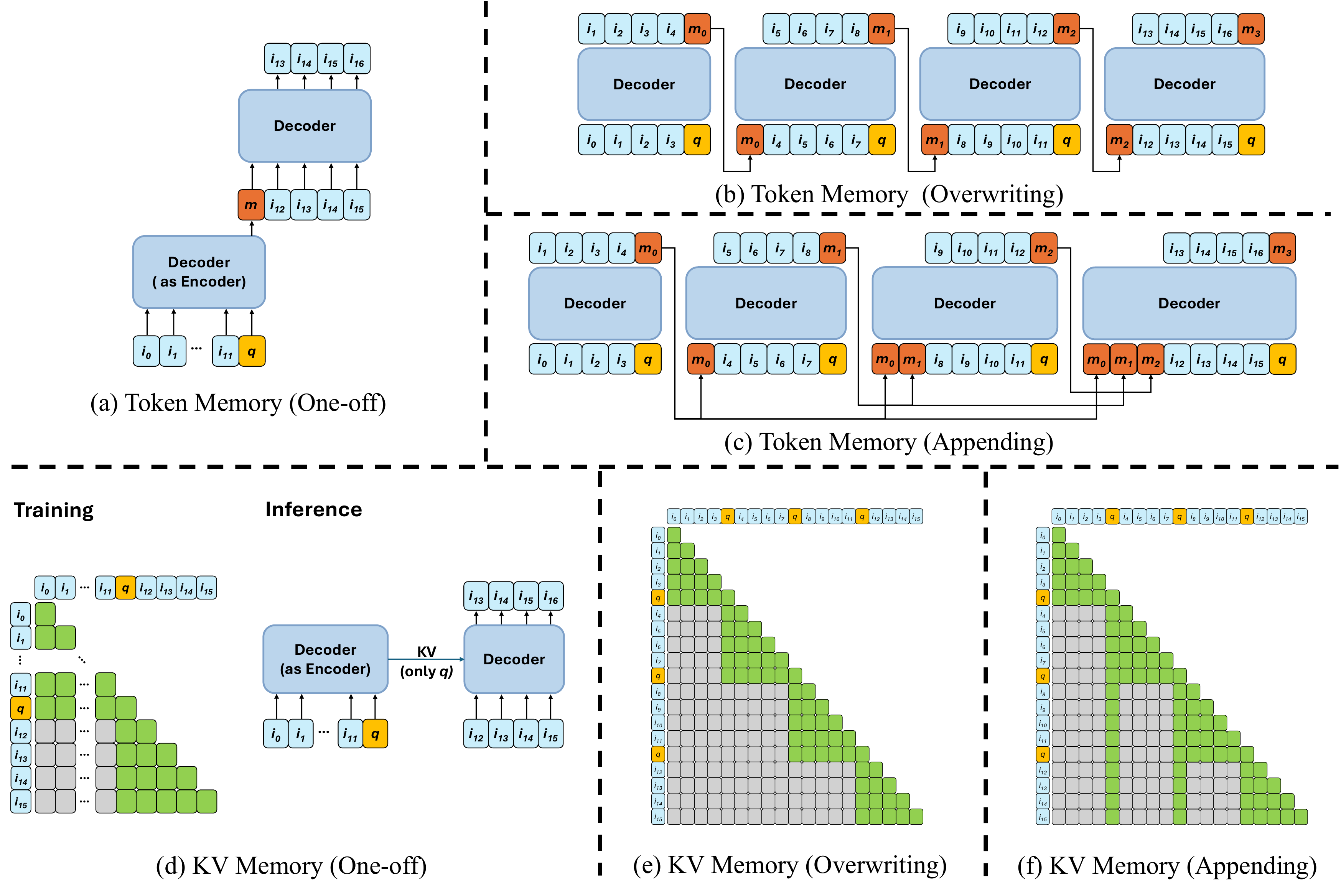} 
    \caption{Taxonomy of Memory Compression Paradigms. Representative methods: (a)~ICAE~\cite{ge2023context}, (b)~RMT~\cite{bulatov2022recurrent}, (c)~AutoCompressors~\cite{chevalier2023adapting}, (d)~Gist~\cite{mu2023learning}, (e)(Hypothetical scheme for comparison), (f)~PersRec~\cite{zhang2026efficient} and Anchor~\cite{pang2024anchor}. For (e) and (f), only the training-time attention mask is shown.}
    \label{fig:arch_comparison}
\end{figure*}

To make long-sequence GenRec practical, we argue that a recommender should \emph{separate what it conditions on by persistence and granularity}, rather than treating the entire raw history as a monolithic context. 

Concretely, we adopt a tripartite memory decomposition (Figure~\ref{fig:CLS}). \textbf{Working Memory} is the recent raw interaction sequence, where high-fidelity details are most important for capturing short-term intent shifts. \textbf{Preference Memory} is a compact, persistent representation extracted from the long archived history, intended to summarize the user's long-term interests. Finally, \textbf{Parametric Memory} refers to the static model weights shared across users, including the encoder/compressor and the autoregressive decoder for next-item prediction.

The core design choice is to compress the massive archived history into compact preference memory that behave like learnable tokens: 
they can be concatenated with Working Memory and processed by standard attention with a cost comparable to short-context modeling. Importantly, this compression also forms an \textit{information bottleneck}: by forcing the model to write history into a limited number of slots, preference memory can filter noisy behaviors and retain preference-relevant signals for prediction.

\textcolor{black}{In large-scale deployment, Preference Memory must be persisted on a per-user basis to avoid repeatedly re-encoding the full interaction history at every step. However, this setting imposes two core system requirements: the memory must support continuous incremental updates as new interactions arrive (Incremental Update), and its representation must remain sufficiently compact to meet the storage constraints of billion-scale user bases (Storage Efficiency). Furthermore, since memory updates at inference time are inherently performed iteratively segment by segment, naively unrolling this recurrent process during training inevitably leads to serial computation graphs, substantially increasing training cost and causing optimization instability. As a result, efficiently training persisted Preference Memory becomes a key challenge.}


\textcolor{black}{Existing solutions in Large Language Models (LLMs) and sequential recommendation struggle to simultaneously satisfy the requirements of incremental updates and storage efficiency for persisted memory. As shown in Figure~\ref{fig:arch_comparison}, storage-friendly token-memory designs often come at the cost of serially unrolled training, while mask-parallel approaches typically rely on persisting per-layer KV caches. To bridge this gap, we propose Rec2PM (Recommendation with Recurrent Preference Memory) for efficient long-sequence generative recommendation. Rec2PM represents Preference Memory as compact compact token embeddings that can be persisted per user and updated iteratively at inference time. More importantly, to avoid serial unrolling during training, Rec2PM introduces a self-referential teacher-forcing objective that enables fully parallel optimization of recurrent updates, while maintaining memory in a storage-efficient token format.}



We further posit that preference memory serve as an effective Information Bottleneck. By compressing history, Rec2PM naturally filters out stochastic noise inherent in long interaction sequences. Consequently, our memory-augmented model captures long-term interests efficiently and, in certain scenarios, can outperform models that directly attend to the full raw sequence. Our contributions are summarized as follows:
\begin{itemize}
    \item We introduce Rec2PM, a generative recommendation framework that compresses long user interaction histories into compact \textit{Preference Memory} tokens. This design enables efficient persistence of users' long-term interests and iterative updates during inference, overcoming the latency bottlenecks of full-attention models while avoiding the high storage costs of KV-cache-based approaches.
    \item We propose a novel self-referential teacher-forcing training strategy that enables fully parallel optimization of recurrent memory updates. This approach leverages global history supervision to avoid the instability and serial computation costs associated with traditional BPTT unrolling.
    \item Experimental results demonstrate that Rec2PM significantly reduces computational and storage costs while achieving superior recommendation accuracy. We further show that Preference Memory serves as an effective denoising \textit{Information Bottleneck}, filtering out stochastic noise from long interaction sequences.
\end{itemize}

\section{Related Works}
\subsection{Generative Recommendation}

The rapid advancement of large language models (LLMs)~\cite{achiam2023gpt,team2023gemini,guo2025deepseek,yang2025qwen3,touvron2023llama} has profoundly impacted sequence modeling across diverse domains. Originally designed for natural language generation, pretrained autoregressive Transformers have demonstrated exceptional generalization capabilities and scalability, prompting their adoption beyond NLP. In the realm of recommender systems, these developments have catalyzed a shift from traditional embedding-based retrieval and multi-stage ranking pipelines~\cite{cheng2016wide,huang2013learning,su2009survey,zhou2018deep} toward unified generative formulations, which reframe recommendation as a sequence generation problem~\cite{zhai2024actions}.

\paragraph{ID-based methods.} Representative approaches~\cite{hidasi2015session,kang2018self,sun2019bert4rec,zhou2020s3} utilize recurrent or self-attention mechanisms to capture temporal dependencies in user interactions. Recently, HSTU~\cite{zhai2024actions} further formulates recommendation as a sequence transduction task within a generative framework, demonstrating significant power-law scaling properties.

\paragraph{LLM-based methods.}
To overcome the limitations of atomic identifiers, ~\cite{rajput2023recommender, agarwal2025pinrec} introduce semantic IDs and multi-token generation strategies. ~\cite{geng2022recommendation, chen2024hllm} reformulate recommendation as a language modeling task. More recently, several studies~\cite{zhou2025onerec,liu2024kuaiformer,qiu2025one,jiang2025large} have proposed unified architectures that integrate retrieval and ranking into end-to-end generative frameworks.

\subsection{Long-Sequence Modeling}
\label{subsection:long-sequence-modeling}

Long-context modeling for LLMs and sequential recommendation increasingly falls under the broader umbrella of \textit{context compression}: reducing the effective context length while preserving task-relevant information~\cite{shi2024keep,zhou2024survey,li2025prompt}.

\paragraph{Hard compression.}
Approaches that directly prune redundant tokens from the context~\cite{li2023compressing,jiang2023llmlingua}. While effective at cutting cost, it is less applicable to recommendation, where identifying redundant historical items without harming preference signals is difficult. 

\paragraph{Soft compression.}
A richer line of work learns continuous latent summaries to replace the original context. Early solutions \cite{dai2019transformer,rae2019compressive,munkhdalai2024leave} extend the receptive field by caching and compressing hidden states across segments, but still require heavy hidden/KV buffers.

More recent methods distill context into compact representations. In LLMs, ICAE~\cite{ge2023context} trains an encoder to map a long prompt into a handful of token embeddings, while 500xCompressor~\cite{li2025500xcompressor} similarly uses an encoder but outputs per-layer KV caches of special tokens. Gist~\cite{mu2023learning} takes a different route: an attention mask distills context into KV caches at designated gist positions in a standard forward pass. In recommendation, KuaiFormer~\cite{liu2024kuaiformer} compresses early- and mid-stage interactions into token embeddings. These methods are mainly one-off compression and do not specify how to update the compressed state under streaming inputs.

RMT~\cite{bulatov2022recurrent} introduces read-write memory tokens that are carried and overwritten across segments, enabling recurrently updates, while AutoCompressors~\cite{chevalier2023adapting} appends newly compressed tokens over time. These token-based memory designs are storage-efficient, but training their update mechanisms typically requires serial unrolling, yielding long back-propagation paths and optimization instability.

Meanwhile, Gist~\cite{mu2023learning} enables fully parallel training through masking, but only supports single-pass compression. Recent works such as PersRec~\cite{zhang2026efficient} (recommendation) and Anchor~\cite{pang2024anchor} (LLMs) segment the sequence in the training mask, allowing segment-wise inference by storing KV caches at anchor positions. However, under our formulation, persisting per-layer KV caches as preference memory---even for a few anchors---is substantially more expensive per user than storing token embeddings.

Beyond sequence compression, some approaches address complementary problems. \cite{li2023prompt} compresses discrete prompts into learned continuous vectors for LLM-based recommendation. \cite{yang2025earn} uses register tokens to compress information within the first $k$ layers and then runs subsequent layers only on these tokens.

\paragraph{Relation to our memory framework.}
Among these methods, those that persist compressed states across inference steps map onto our memory framework. Figure~\ref{fig:arch_comparison} summarizes representative designs by persisted format (token embeddings vs.\ KV caches) and update mechanism (one-off vs.\ overwriting updates vs.\ appending updates). Compared with these memory schemes, Rec2PM supports parallel training and recurrent updates at inference time, and persists lightweight token embeddings as preference memory rather than KV caches.
\section{Methodology}
In this section, we present \textbf{Rec2PM}, our recurrent preference memory framework for generative recommendation. We first formulate the problem and the general memory-augmented setting. Next, we detail our architecture, focusing on the learnable atomic memory states and the associated encoding and updating mechanisms. Finally, we introduce our self-referential training strategy, which enables efficient parallel training while maintaining the integrity of recurrent memory updates.

\subsection{Problem Formulation}

\subsubsection{Autoregressive Modeling for Sequential Recommendation}

Let $\mathcal{I}$ denote the set of items. For each user, the interaction history is represented as a sequence $S = \{I_0, I_1, \dots, I_{n-1}\}$, where $I_i \in \mathcal{I}$ is the item interacted with at time step $i$. The goal of a generative recommender is to model the probability distribution of the next item $I_n$ conditioned on the historical sequence $S$:

\begin{equation}
    P(I_n | S) = P(I_n | I_0, I_1, \dots, I_{n-1}; \theta)
\end{equation}

where $\theta$ denotes the model parameters.

For autoregressive modeling over the entire sequence, we maximize the likelihood of each next interaction conditioned on its prefix. Concretely, given a minibatch of $B$ sequences, we minimize the negative log-likelihood:

\begin{equation}
    \mathcal{L}_{AR} = - \sum_{j=0}^{B-1} \sum_{i=1}^{n-1} \log P(I_{j,i} \mid I_{j,0:i-1}; \theta),
\end{equation}

where $n$ is the sequence length, and $I_{j,0:i-1}$ denotes the prefix subsequence $\{I_{j,0}, \dots, I_{j,i-1}\}$.

This probability is typically modeled using a Transformer-based architecture. To strictly enforce the autoregressive property, a causal attention mask is applied within the self-attention mechanism.

\subsubsection{Preference Memory for Lifelong Sequences}

In real-world recommendation scenarios, user interaction sequences can be extremely long. To manage this complexity, we adopt a memory-augmented approach where historical information is maintained in a compressed preference memory.

We segment the user's history sequence $S$ into fixed-length segments. Let $L_{seg}$ denote the length of each segment. The sequence $S$ is partitioned into $S = \{S_0, S_1, \dots, S_k\}$, where each completed segment $S_j$ (for $j < k$) contains exactly $L_{seg}$ interactions, and the final segment $S_k$ contains the remaining interactions ($|S_k| \le L_{seg}$).

Formally, when predicting the item at step $i$, the user's history is divided into two parts:
\begin{itemize}
    \item $S_{hist}$: all completed segments before the current one($S_{hist} = \{S_0, \dots, S_{k-1}\}$). We denote $M_{k-1}$ as the preference memory that compresses information from these historical segments.
    \item $S_{recent}$: the current segment $S_k$, which contains the most recent interactions that have not yet been compressed.
\end{itemize}

The prediction of the next item $I_i$ depends on both the compressed preference memory of the past and the detailed recent context:
\begin{equation}
    P(I_i | S) \approx P(I_i | M_{k-1}, S_k)
\end{equation}

When the current segment $S_k$ reaches the full length $L_{seg}$, a memory update is triggered. The content of $S_k$ is compressed and merged into the preference memory, transitioning it from $M_{k-1}$ to $M_k$.

\subsection{Learnable Tokens as Preference Memory}
\label{subsection:learnable-tokens-as-memory-states}

In this work, we introduce \textbf{Atomic Memory State} $m$ to explicitly capture long-term user interests. This atomic memory state is derived by interacting a set of global learnable parameters with the user's input context. We denote these parameters as \textbf{Memory Queries Vectors}, $Q_{mem} \in \mathbb{R}^{C \times d}$, where $C$ represents the number of memory slots and $d$ is the embedding dimension.

We define a generic memory encoding operation that transforms an input context into a compressed preference memory. By concatenating the input context with the memory queries and feeding them into a Memory Encoder, we extract the token embeddings corresponding to the query positions as the resulting atomic memory state $m$:

\begin{equation}
    \begin{aligned}
    H &= \text{Encoder}([E_{encode}; Q_{mem}]) \\
    m &= H_{|E_{encode}|+1:|E_{encode}|+C}
    \end{aligned}
\end{equation}

Here, $E_{encode}$ represents the source information to be compressed. Consequently, $m$ and $Q_{mem}$ share the same shape $\mathbb{R}^{C \times d}$. Crucially, $Q_{mem}$ functions as a memory extractor learned globally during training (Parametric Memory), whereas $m$ models personalized long-term interests for a specific user during inference (Preference Memory).

The proposed memory mechanism operates in two phases: initialization (cold start) and recurrent update (streaming). We introduce two update variants: \textbf{Overwriting} and \textbf{Appending}. Let $m_k$ denote the atomic memory generated at step $k$, and $M_k$ denote the effective memory context available after step $k$.
\begin{itemize}
    \item \textbf{Overwriting Mode:} Overwriting the previous preference memory with the most recent atomic memory state: $M_k = m_k$. This yields a constant-size preference memory.
    \item \textbf{Appending Mode:} Appending the new atomic memory to the existing preference memory: $M_k = [M_{k-1}; m_k]$. This yields a growable preference memory.
\end{itemize}

\paragraph{Initialization.}
When a user sequence begins, the first segment $S_0$ serves as the initial context. We treat $S_0$ as the input $E_{encode}$ in the encoding operation. 
\begin{equation}
    E_{encode, 0} = S_0
\end{equation}
$Q_{mem}$ interact with the raw items in $S_0$ via the encoder to generate the initial atomic memory state $m_0$. Ideally, $M_0 = m_0$ for both modes.

\paragraph{Recurrent Update.}
As the user interacts with more items, the system maintains a working context $S_{recent}$ (corresponding to the current segment $S_k$). Once $S_k$ reaches the predefined segment length $L_{seg}$, a memory update is triggered. To capture both long-term history and recent dynamics, we construct the input context $E_{encode, k}$ by concatenating the previous preference memory $M_{k-1}$ with the newly completed segment $S_k$:
\begin{equation}
    E_{encode, k} = [M_{k-1}; S_k]
\end{equation}
We then apply the encoding operation to obtain the atomic memory state $m_k$. We subsequently update $M_k$ according to the chosen mode (Overwriting or Appending).

\begin{figure*}[htbp]
    \centering
    \includegraphics[width=0.9\linewidth]{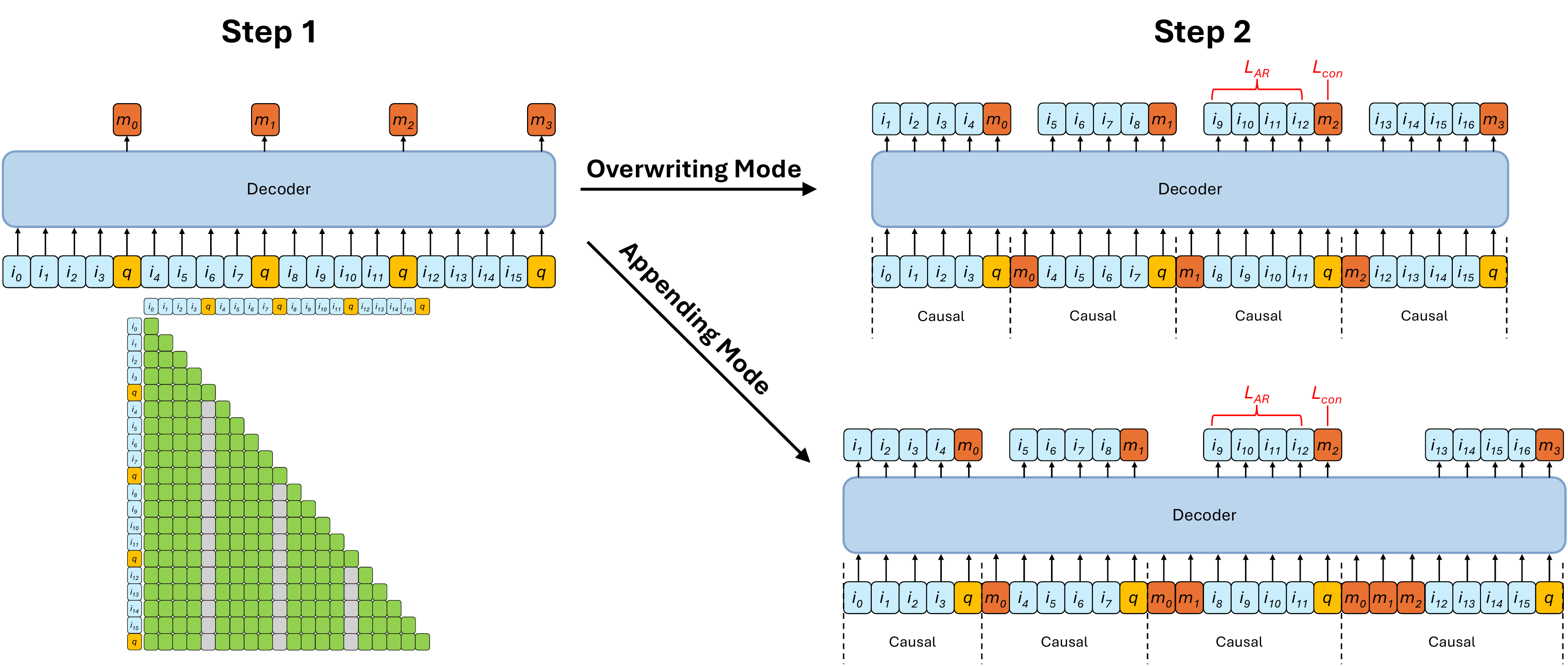}
    \caption{Illustration of the proposed two-stage parallel training paradigm. Stage 1 generates global reference memories by attending to raw history. Stage 2 performs parallel optimization for incremental updates (in Overwriting or Appending modes) under the supervision of the reference memories.}
    \label{fig:training_paradigm}
\end{figure*}

\paragraph{Memory Utilization for Prediction.}
To predict the next item within the current segment $S_k$, the Decoder requires access to both the compressed preference memory and the detailed recent interactions. We construct the decoder input by concatenating them:
\begin{equation}
    E_{decode, k} = [M_{k-1}; S_k]
\end{equation}
The decoder processes this sequence via causal self-attention, where each position in $S_k$ attends to $M_{k-1}$ and preceding tokens in $S_k$ to predict the next item.

\paragraph{Unified Architecture.}
Although memory updating and item prediction are conceptually distinct operations (one generates $m_k$, the other generates predictions for $S_k$), they share the same input context structure. We therefore implement them within a single unified architecture where the Memory Encoder and Generative Decoder share parameters. To maximize efficiency, we perform both tasks in a single forward pass by constructing a joint sequence:
\begin{equation}
    E_{unified, k} = [M_{k-1}; S_k; Q_{mem}]
\end{equation}
Under standard causal masking, the tokens in $S_k$ attend to $M_{k-1}$ and their predecessors to predict the next items, while the tokens in $Q_{mem}$ (positioned at the end) naturally attend to both $M_{k-1}$ and the full $S_k$ to generate the updated atomic memory state $m_k$. This unified design ensures that memory representations are directly optimized for the prediction objective and reduces the computational overhead of memory maintenance.

\subsection{Training}
\label{section:training}

While our Preference Memory architecture inherently supports recurrent updates, training this mechanism sequentially presents significant challenges. Figures~\ref{fig:arch_comparison}(b) and (c) illustrate the serially unrolled token-memory paradigm typically employed in recurrent memory models~\cite{bulatov2022recurrent, chevalier2023adapting}.

Training in this sequential manner faces two critical issues. First, it necessitates maintaining long gradient chains across multiple steps for Back-Propagation Through Time (BPTT), resulting in prohibitive computational overhead. Second, even when mitigating computational costs with techniques like stop-gradients~\cite{chevalier2023adapting}, sequential training remains vulnerable to error accumulation. This instability complicates optimization and often leads to suboptimal convergence.

Mask-parallel approaches \cite{mu2023learning, zhang2026efficient}, as shown in Figure~\ref{fig:arch_comparison}(d-f), rely on caching Key-Value pairs for all history layers as. While effective for parallelism, this results in a memory footprint significantly larger than our compact token embeddings, violating our storage constraints. To reconcile the conflict between efficient parallel training and compact memory states, we propose a self-referential training strategy inspired by teacher forcing, as illustrated in Figure~\ref{fig:training_paradigm}. In this paradigm, the model generates its own ``teacher" signals by attending to the global history, using them to supervise the updates.

\subsubsection{Parallel Training via Self-Referential Teacher Forcing}
Given a user sequence partitioned into segments $S = \{S_0, S_1, \dots, S_k\}$, we process the data in two passes:

\paragraph{Step 1: Global Reference Generation.}
First, we construct a global ``reference" memory by allowing the model to view the raw history directly. We interleave $Q_{mem}$ into the full sequence of segments:
\begin{equation}
    E_{global} = [S_0; Q_{mem}; S_1; Q_{mem}; \dots; S_k; Q_{mem}]
\end{equation}
We apply a customized attention mask to $E_{global}$ where tokens can attend to their causal history of raw items ($S_{0:i}$) but are prevented from attending to preceding $Q_{mem}$ tokens. Consequently, the output of $Q_{mem}$ at the end of segment $S_h$ is equivalent to compressing the entire raw prefix $[S_0, \dots, S_h]$ directly. We denote this output as the \textbf{Reference Memory}, $m_h^{(ref)}$. Since $m_h^{(ref)}$ is derived from the full raw history, we consider it a low-error atomic memory state. Also, we denote $M_h^{(ref)} = m_h^{(ref)}$ for Overwriting and $M_h^{(ref)} = [m_0^{(ref)}; m_1^{(ref)}; \dots; m_h^{(ref)}]$ for Appending.
\paragraph{Step 2: Parallel Prediction \& Update.}
In the second step, we simulate the recurrent update process in parallel. We divide the sequence into independent subsequences. For each segment $S_h$, we construct the input by combining ($M_{h-1}^{(ref)}$) from the previous step with the current segment:
\begin{equation}
    E_{local, h} = [M_{h-1}^{(ref)}; S_h; Q_{mem}]
\end{equation}
These subsequences are processed in parallel by the Decoder. Within each subsequence:
\begin{itemize}
    \item The tokens in $S_h$ perform causal attention to predict the next items, supervised by the autoregressive loss $\mathcal{L}_{AR}$.
    \item The $Q_{mem}$ tokens attend to $M_{h-1}^{(ref)}$ and $S_h$ to generate the \textbf{Updated Memory}, $m_h^{(upd)}$.
\end{itemize}

\paragraph{Optimization Objective.}
The updated memory $m_h^{(upd)}$ represents the state obtained via a single update step. To ensure the recurrent memory mechanism remains stable over long sequences, we enforce consistency between the updated memory and the reference memory. We minimize the Mean Squared Error (MSE) between them:
\begin{equation}
    \mathcal{L}_{con} = \frac{1}{k} \sum_{h=0}^{k} \| m_h^{(ref)} - m_h^{(upd)} \|^2
\end{equation}
The final training objective combines the recommendation task with the memory consistency constraint:
\begin{equation}
    \mathcal{L} = \mathcal{L}_{AR} + \lambda \mathcal{L}_{con}
\end{equation}
where $\lambda$ is a hyperparameter balancing the two terms. This approach allows us to train on all segments in parallel while ensuring the preference memory effectively captures the global history.

\subsubsection{Justification of the Training Paradigm.}
The effectiveness of this strategy stems from the synergy between the two objectives and the teacher-forcing nature of the architecture. We articulate the rationale as follows:

\paragraph{Implicit Supervision for High-Quality Compression.} The autoregressive loss $\mathcal{L}_{AR}$ acts on items conditioned on $M_{h-1}^{(ref)}$. To minimize prediction error, the model is compelled during Stage 1 to compress all essential information from the raw prefix into the reference memory. This ensures that our ``Teacher'' signal is semantically rich and accurate. 

\paragraph{Supervision for Recurrent Updates.} The consistency loss $\mathcal{L}_{con}$ explicitly ensures that the update operation mimics the global compression. It forces the model to learn how to transition from $M_{h-1}$ to $M_h$ without losing information, effectively transferring the compression capability of the global view to the recurrent updater.

\paragraph{Stabilization via Teacher Forcing.} Crucially, during Stage 2, the generation of the updated memory $m_h^{(upd)}$ is conditioned on the high-quality reference $M_{h-1}^{(ref)}$, rather than a rolled-out state with accumulated errors. This decouples the training steps and prevents the ``drift'' phenomenon often seen in RNN training, effectively applying Teacher Forcing to the memory mechanism for stable and efficient convergence.

\subsubsection{Discussion: Implicit Supervision vs. Reconstruction.}
\label{subsection:implicit-supervision-vs-reconstruction}
We deliberately exclude an explicit reconstruction loss (e.g., forcing preference memory to reconstruct raw history~\cite{ge2023context, li2025500xcompressor}). From an Information Bottleneck~\cite{tishby2000information} perspective, our goal is to maximize predictive information $I(M; Y)$ subject to a capacity constraint imposed by the limited memory tokens. Explicit reconstruction forces the maximization of $I(M; S_{hist})$, compelling the memory to encode high-entropy noise (e.g., random clicks) alongside valid interests. Given the tight memory bottleneck, this leads to \textit{capacity contention}, where noise displaces predictive signals. By relying solely on the implicit supervision from the autoregressive task ($\mathcal{L}_{AR}$), the Decoder acts as a critic, guiding the memory to discard noise and retain only the latent intent necessary for future prediction.

\section{Experiments}
\subsection{Main Experiments Settings}
\subsubsection{Datasets and preprocessing.}

We conduct our experiments on the MerRec dataset~\cite{merrec}, a large-scale real-world benchmark collected from the Mercari C2C platform. Crucially for our study, unlike many traditional datasets dominated by short sessions, MerRec contains a significant proportion of extremely long user interaction sequences. This makes it an ideal testbed for evaluating the capability of recommendation models to capture long-term dependencies from extensive historical contexts.

To strictly evaluate long-sequence modeling capabilities, we filter for users with at least $1003$ interactions. For each user, we truncate the interaction sequence to the most recent $1003$ items. We adopt a leave-one-out evaluation strategy: for each user's sequence, the second-to-last interaction is reserved for validation, the last interaction is used for testing, and the preceding interactions form the training set.

\subsubsection{Context length settings.}
To rigorously evaluate the impact of context length, we define two input settings for backbone baselines without memory:
\begin{itemize}
    \item \textit{Short}: We set $L_{\text{short}}{=}200$. To prevent data wastage during training, we partition the long training prefix (length $L_{\text{full}}$) into non-overlapping chunks of length $L_{\text{short}}$, treating each chunk as an independent training instance. During evaluation, only the most recent $L_{\text{short}}$ interactions are used.
    \item \textit{Full}: We set $L_{\text{full}}{=}1000$. We feed the entire training prefix into the model and optimize the standard autoregressive next-item prediction objective ($\mathcal{L}_{AR}$) over the sequence.
\end{itemize}

For all memory-augmented variants, we maintain an input length of $L_{\text{full}}$ but process the sequence in segments of length $L_{\text{seg}}{=}200$. This results in $L_{\text{full}}/L_{\text{seg}}$ segments per user. During inference, the model updates the preference memory segment-by-segment and generates predictions based on the output of the final segment.

\begin{table*}[ht]
    \caption{Main results on the MerRec long-sequence benchmark. We compare our Rec2PM against token-memory baselines with serially unrolled training (Tok-Serial-*) and KV-cache baselines with mask-parallel training (KV-Mask-*) on two backbones (SASRec and HSTU). The best results are in bold and the second best results are underlined.}
    \label{tab:main_results_merrec}
    \begin{tabular}{lccccc ccccc}
      \toprule
      \multirow{2}{*}{\textbf{Method}} & \multicolumn{5}{c}{\textbf{SASRec ~\cite{kang2018self}}} & \multicolumn{5}{c}{\textbf{HSTU ~\cite{zhai2024actions}}} \\
      \cmidrule(lr){2-6}\cmidrule(lr){7-11}
       & \textbf{H@1} & \textbf{H@10} & \textbf{H@50} & \textbf{N@10} & \textbf{N@50} & \textbf{H@1} & \textbf{H@10} & \textbf{H@50} & \textbf{N@10} & \textbf{N@50} \\
      \midrule
      Short & 14.10 & 40.96 & 57.59 & 26.68 & 30.39 & 13.94 & 41.67 & 59.08 & 26.86 & 28.88 \\
      Full & 14.43 & 42.40 & 59.31 & 27.44 & 31.23 & 14.24 & 42.77 & 60.37 & 27.47 & 31.41 \\
      Tok-Serial-O & 14.57 & 42.56 & 59.66 & 27.62 & 31.46 & 14.65 & 43.75 & 61.03 & 28.20 & 32.07 \\
      Tok-Serial-A & 14.49 & 42.56 & 59.70 & 27.58 & 31.43 & 14.45 & 43.60 & 61.02 & 28.01 & 31.91 \\
      KV-Mask-O & \underline{14.73} & 42.32 & 59.31 & 27.60 & 31.41 & 14.56 & 43.64 & 61.07 & 28.08 & 32.00 \\
      KV-Mask-A & 14.72 & 42.35 & 59.37 & 27.56 & 31.37 & 14.64 & 43.59 & 60.88 & 28.10 & 31.97 \\
      Rec2PM-O & \textbf{14.79} & \textbf{43.12} & \textbf{59.92} & \textbf{28.05} & \textbf{31.82} & \textbf{15.04} & \textbf{44.20} & \textbf{61.23} & \textbf{28.66} & \textbf{32.48} \\
      Rec2PM-A & \underline{14.73} & \underline{42.76} & \underline{59.74} & \underline{27.81} & \underline{31.62} & \underline{14.87} & \underline{44.13} & \underline{61.16} & \underline{28.50} & \underline{32.31} \\
    \bottomrule
    \end{tabular}
\end{table*}

\subsubsection{Backbones and compared methods.}
We instantiate memory mechanisms on top of two representative sequential recommendation backbones: SASRec~\cite{kang2018self} and HSTU~\cite{zhai2024actions}. The comparison includes two baselines without persisted preference memory and six memory-augmented variants:

\begin{itemize}
    \item \textit{Short/Full}: Baselines without preference memory.
    \item \textit{Tok-Serial-O/A}: token-memory with serially unrolled training (Figure~\ref{fig:arch_comparison}(b,c))~\cite{bulatov2022recurrent,chevalier2023adapting}. We evaluate both overwriting (O) and appending (A) updates at inference.
    \item \textit{KV-Mask-O/A}: KV-cache memory with mask-parallel training (Figure~\ref{fig:arch_comparison}(e,f))~\cite{mu2023learning,zhang2026efficient,pang2024anchor}. The attention mask forces cross-segment information flow to go through designated memory positions; we evaluate both overwriting (O) and appending (A) updates at inference.
    \item \textit{Rec2PM-O/A} (ours): token-memory with parallel training via our proposed self-referential teacher-forcing objective. At inference time, Rec2PM follows the same overwriting(O) and appending(A) update rules as Tok-Serial.
\end{itemize}

\subsubsection{Hyperparameters and metrics.}

We set the embedding dimension to $64$ for all models. For SASRec, we employ $4$ layers with $4$ attention heads, while for HSTU, we use $16$ layers with $8$ heads. Models are trained with a learning rate of $10^{-3}$, batch size of $8$, and a weight decay of $0.1$, using an early-stopping patience of $10$ epochs. The memory slots number is set to $C{=}4$. For our proposed method, the consistency loss weight is set to $\lambda{=}1$. 

Performance is evaluated using Hit Rate (H@K) and NDCG (N@K). For all experiments conducted on the MerRec dataset, we report the average results over five independent runs using fixed random seeds ranging from 0 to 4.

\subsection{Main results}
\label{subsection:main-results}

Based on the results in Table~\ref{tab:main_results_merrec}, we make the following observations:

\paragraph{Memory as a Denoiser.} 
Both memory-augmented models and the Full baseline significantly outperform the Short baseline, confirming that user history contains valuable information for predicting future interactions. Notably, despite having a much smaller effective context window than Full, memory-based models achieve comparable or even superior performance. We attribute this to the noisy nature of user interaction data in recommendation systems. Direct attention over extremely long sequences (as in Full) is prone to distraction by irrelevant stochastic behaviors. As discussed in Section~\ref{subsection:implicit-supervision-vs-reconstruction}, the Preference Memory acts as an Information Bottleneck, compressing history into limited slots. This forces the model to distill only the most salient semantic information while filtering out noise, often yielding better generalization than attending to the raw long sequence.

\paragraph{Ineffectiveness of Appending.} 
Across all variants, appending schemes do not improve performance over overwriting schemes and often degrade it slightly. This further supports the noise hypothesis: overwriting updates compel the model to discard less relevant information to make room for new updates, maintaining a strict bottleneck. In contrast, appending updates accumulate historical states, weakening the bottleneck effect and retaining more noise. Furthermore, unlike in LLMs where specific retrieval from distant history is crucial, sequential recommendation is dominated by recency effects~\cite{liu2024kuaiformer}. Appending memory may distract the attention mechanism with stale history, reducing the focus on more critical recent interactions.

\paragraph{Superiority of Rec2PM.} 
Finally, our Rec2PM-O scheme consistently outperforms other memory baselines. Compared to token-memory with serially unrolled training (Tok-Serial), Rec2PM benefits from our parallel training paradigm with Teacher Forcing, which mitigates the error accumulation inherent in serial training. Compared to KV-cache memory with mask-parallel training (KV-Mask), Rec2PM offers better information flow. In KV-Mask schemes, information from previous segments is restricted by the attention mask and may not be fully propagated to the Query Tokens' KV cache at lower layers. In contrast, Rec2PM explicitly provides the mature, pre-computed token embeddings of the previous segment at the input layer, allowing the current segment to attend to the full historical context from the very first transformer layer.

\subsection{In-Depth Analysis}

\subsubsection{One-time Full-sequence Compression}

\begin{table}[h]
  \caption{Performance comparison of the trained Rec2PM-O model under two inference protocols: standard \textit{Iterative} updates vs. \textit{One-off} compression. In the One-off setting, we compress all segments (except the last one) into the preference memory in a single step, which is then concatenated with the final segment for prediction. Results are reported on MerRec with the HSTU backbone.}
  \label{tab:onetime_compression_hstu_merrec}
  \begin{tabular}{lcccc}
    \toprule
     & \textbf{H@1} & \textbf{H@10} & \textbf{H@50} & \textbf{N@50} \\
    \midrule
    Iterative & 15.04 & 44.20 & 61.23  & 32.48 \\
    One-off & 15.05 & 44.20 & 61.23  & 32.48 \\
    \bottomrule
  \end{tabular}
\end{table}

As discussed in Section~\ref{section:training}, our proposed parallel training strategy trains the model to perform incremental memory updates while aligning them with a global reference derived from the raw history. This implies that the model should theoretically support both iterative updates and one-time global compression. To verify this, we directly evaluate the trained Rec2PM-O model from Section~\ref{subsection:main-results} (without fine-tuning) under a \textbf{One-time} compression setting: we compress the entire prefix sequence into the preference memory in a single forward pass and use it to predict items in the final segment.

The results, presented in Table~\ref{tab:onetime_compression_hstu_merrec}, show that the one-time compression yields performance remarkably consistent with the standard iterative inference. This demonstrates the robustness of our memory mechanism and validates the effectiveness of the consistency loss in aligning the incremental state with the global history representation.

\subsubsection{Ablation on Consistency Loss}

\begin{table}[h]
    \caption{Ablation study on the consistency loss ($\mathcal{L}_{con}$). We compare the default Rec2PM-O (with $\lambda{=}1$) against a variant trained without consistency loss ($\lambda{=}0$). Results are reported on MerRec with the HSTU backbone.}
    \label{tab:ablation_consistency_loss}
    \begin{tabular}{lcccc}
      \toprule
       & \textbf{H@1} & \textbf{H@10} & \textbf{H@50} & \textbf{N@50} \\
      \midrule
      Rec2PM-O & 15.04 & 44.20 & 61.23  & 32.48 \\
      Rec2PM-O w/o $\mathcal{L}_{con}$ & 14.43 & 43.38 & 61.20 & 31.86 \\
      \bottomrule
    \end{tabular}
  \end{table}

To further demonstrate the effectiveness of the Consistency Loss, we removed $\mathcal{L}_{con}$ from the training of Rec2PM-O, retaining only $\mathcal{L}_{AR}$ (setting $\lambda{=}0$). 

The results shown in Table~\ref{tab:ablation_consistency_loss} indicate a performance drop when the consistency loss is removed. This degradation occurs because, without $\mathcal{L}_{con}$, the model only receives supervision for the one-time global compression (propagated via $\mathcal{L}_{AR}$ to Step 1) during training, leaving the update mechanism in Step 2 unsupervised. Consequently, the model fails to learn the transition from $M_{h-1}$ to $M_{h}$. This discrepancy between the training objective and the iterative inference requirement leads to poor performance, thereby highlighting the necessity of the consistency loss.

\subsubsection{Impact of Memory Slots Number}

\begin{table}[h]
    \caption{Impact of memory slots number on the performance of Rec2PM-O. We compare the performance of Rec2PM-O with different number of memory slots $C=\{1, 2, 4, 8, 16\}$ on MerRec with the HSTU backbone.}
    \label{tab:impact_of_memory_slots}
    \begin{tabular}{lcccc}
      \toprule
       & \textbf{H@1} & \textbf{H@10} & \textbf{H@50} & \textbf{N@50} \\
      \midrule
      Full & 14.24 & 42.77 & 60.37 & 31.41 \\
      Rec2PM-O $C=1$ & 15.05 & 44.14 & 60.92 & 32.42 \\
      Rec2PM-O $C=2$ & 14.89 & 44.38 & 61.22 & 32.44 \\
      Rec2PM-O $C=4$ & 15.04 & 44.20 & 61.23 & 32.48 \\
      Rec2PM-O $C=8$ & 14.80 & 44.18 & 61.45 & 32.36 \\
      Rec2PM-O $C=16$ & 14.72 & 44.10 & 61.19 & 32.30 \\
      \bottomrule
    \end{tabular}
  \end{table}

We explore the impact of the number of memory slots on performance. As shown in Table~\ref{tab:impact_of_memory_slots}, overall, our method exhibits stable performance across different numbers of memory slots and consistently outperforms the Full baseline on the MerRec dataset. One observation is that performance degrades slightly when the number of memory slots is extremely small ($C=1$) or extremely large ($C=16$). We hypothesize that when the number of slots is too small, the capacity is insufficient to adequately maintain preference information from the historical interaction sequence. Conversely, when the number of slots is too large, the Memory fails to serve as an effective information bottleneck, introducing noise that interferes with attention.

\subsubsection{Efficiency Analysis}

We evaluate the computational and storage efficiency on MerRec with the HSTU backbone (Short: 200, Full: 1000; memory variants: $L_{\text{seg}}{=}200$ with the first four segments compressed and persisted as Preference Memory). Table~\ref{tab:efficiency_analysis} shows that memory-based variants achieve latency comparable to Short ($\sim$10ms) while Full is substantially slower (135ms). Moreover, Rec2PM is far more storage-efficient than KV-Mask methods.

\begin{table}[t]
  \caption{Efficiency comparison on MerRec (HSTU). The Storage column reports the per-user preference memory footprint, and the Latency column reports the model-internal inference time on an NVIDIA H20 GPU with batch size 64.}
  \label{tab:efficiency_analysis}
  \begin{tabular}{lcc}
    \toprule
    \textbf{Method} & \textbf{Storage} & \textbf{Latency} \\
    \midrule
    Short & -- & 9ms \\
    Full & -- & 135ms \\
    KV-Mask-O & 32KB & 10ms \\
    KV-Mask-A & 128KB & 10ms \\
    Rec2PM-O & 1KB & 10ms \\
    Rec2PM-A & 4KB & 10ms \\
    \bottomrule
  \end{tabular}
\end{table}

\subsection{Evaluation on Industrial Dataset}
We further validate our method on a large-scale proprietary dataset collected from a major commercial content recommendation service. More details are provided in Appendix~\ref{appendix:industrial_dataset_stat}. 
We follow a realistic chronological split (train on multiple weeks of logs and evaluate on the subsequent day).
We use \textit{HSTU-Full} with a context length of 2048 as the primary full-context baseline, along with shorter-context variants with different input lengths. For our Rec2PM method, we perform only one-time compression: we compress the long history (the first 1948 interactions) into $C{=}20$ memory slots, and concatenate them with the most recent 100 interactions for next-item prediction.

\begin{table}[t]
  \caption{Main results on the industrial dataset. The best results are in bold and the second best results are underlined.}
  \label{tab:industrial_main_results}
  \centering
  \begin{tabular}{lcccc}
    \toprule
    \textbf{Model}  & \textbf{H@1} & \textbf{H@10} & \textbf{H@50} & \textbf{H@1000} \\
    \midrule
    \textbf{HSTU-Full}  & \textbf{0.21} & 1.88 & 6.18 & 31.81 \\
    \midrule
    \quad Seq-50   & 0.17 & 1.89 & 5.83 & 30.88 \\
    \quad Seq-100   & 0.18 & 1.84 & 6.04 & 31.17 \\
    \quad Seq-200   & \underline{0.20} & 1.95 & 6.20 & 31.42 \\
    \quad Seq-500   & 0.19 & \textbf{2.12} & \underline{6.55} & \underline{32.32} \\
    \quad Seq-1000   & \underline{0.20} & 1.95 & 6.25 & 31.97 \\
    \midrule
    \textbf{Rec2PM} &  \textbf{0.21} & \underline{2.08} & \textbf{6.64} & \textbf{33.06} \\
    \bottomrule
    \end{tabular}
  \end{table}

As shown in Table~\ref{tab:industrial_main_results}, the memory-based approach (Rec2PM) outperforms the full-sequence attention baseline (HSTU-Full). Moreover, without a Preference Memory, simply increasing the raw context length yields diminishing returns and can even slightly degrade performance (e.g., Seq-500 vs. Seq-1000). Overall, these findings again corroborate our earlier analysis: the Preference Memory serves as an information bottleneck that distills salient user interests and filters noise, whereas attending to longer uncompressed histories may amplify noise rather than add useful signal.

\section{Conclusion}
\label{section:conclusion}
In this paper, we introduced \textbf{Rec2PM}, a scalable framework designed to overcome the computational bottlenecks in long-sequence generative recommendation. By compressing extensive user histories into compact \textit{Preference Memory} tokens, Rec2PM enables efficient storage of long-term interests without extensive KV caches. A key innovation of our work is the self-referential teacher-forcing training paradigm, which successfully bridges the gap between parallelizable training and recurrent inference, allowing the model to learn high-quality memory updates without the instability and inefficiency of serial unrolling. Empirical results validate that Rec2PM significantly outperforms existing baselines in accuracy while reducing the per-user memory footprint by orders of magnitude. Furthermore, our analysis confirms that the Preference Memory functions as a critical \textit{Information Bottleneck}, effectively denoising stochastic user behaviors to retain only the most predictive signals. We believe Rec2PM provides a robust foundation for deploying lifelong user modeling in real-world recommendation systems.

\bibliographystyle{ACM-Reference-Format}
\bibliography{sample-base}

\appendix

\section{Experiments on Industrial Dataset}
\label{app:industrial_exp}

We conducted experiments on a large-scale industrial dataset to evaluate the scheme of compressing user interaction sequences into Preference Memory. In this section, we provide a detailed description of the industrial dataset and present additional experimental results.

\subsection{Dataset Statistics}
\label{appendix:industrial_dataset_stat}

We conduct experiments on a large-scale industrial dataset collected from a short-video platform. To reflect the real-world data distribution, we use logs from 2025/04/01 to 2025/05/16 for training and 2025/05/17 for testing. The raw data contains over 500 billion interactions. Detailed statistics are summarized in Table~\ref{tab:industrial_stats}.

\begin{table}[h]
    \caption{Statistics of the industrial dataset.}
    \label{tab:industrial_stats}
    \centering
    \begin{tabular}{lr}
      \toprule
      \textbf{Statistic} & \textbf{Value} \\
      \midrule
      Users & $\approx$ 500 Million \\
      Items &  $>$ 550 Million \\
      Interactions & $>$ 500 Billion \\
      \midrule
      Time Range (Train) & 2025/04/01 -- 2025/05/16 \\
      Time Range (Test) & 2025/05/17 \\
      \midrule
      Avg. Sequence Length & 1,147 \\
      Max. Sequence Length & 2,048 \\
      Data Size (Disk) & $\approx$ 200 TB \\
      \bottomrule
    \end{tabular}
\end{table}

\subsection{Experimental Settings}
We use the HSTU backbone with a context length of 2048 as the primary full-context baseline, along with shorter-context variants with different input lengths. For our preference memory mechanism, we perform only one-time compression: we compress the long history (the first 1948 interactions) into $C{=}20$ memory slots, and concatenate them with the most recent 100 interactions for next-item prediction.

\subsection{Implicit vs. Explicit Supervision}
We verify our core theoretical claim in Section~\ref{subsection:implicit-supervision-vs-reconstruction}: that explicit reconstruction is harmful for the model to fully store useful user long-term preferences in the limited memory. 

\begin{table}[h]
\centering
\caption{Impact of Supervision Signals (HR@1000). Adding reconstruction loss degrades performance, supporting the Information Bottleneck principle.}
\label{tab:ablation_supervision}
\begin{tabular}{lcc}
\toprule
\textbf{Variant} & \textbf{HR@1000} & \textbf{Rel. Change} \\
\midrule
Implicit Supervision Only & \textbf{0.3306} & - \\
\midrule
Adding $\mathcal{L}_{Rec}$ & 0.3243 & -1.9\% \\
\bottomrule
\end{tabular}
\end{table}

Specifically, we compare our Implicit Supervision strategy against a variant trained with explicit reconstruction objectives. As shown in Table~\ref{tab:ablation_supervision}, adding the explicit Reconstruction Loss ($\mathcal{L}_{Rec}$) degrades performance (-1.9\%). 

This validates the Capacity Contention theory: forcing the limited memory to reconstruct raw history compels it to waste capacity on memorizing noise. By relying solely on implicit supervision, the memory is free to ignore irrelevant details and focus purely on predictive patterns.

\subsection{Robustness to Temporal Overlap}
We also evaluate the model's robustness to temporal inconsistency. In practical streaming engineering, due to log delays, there is often an overlap between the history compressed in Memory and the recent sequence input to the model. We test this scenario in Table~\ref{tab:ablation_overlap}.

\begin{table}[h]
\centering
\caption{Robustness to Temporal Overlap (HR@1000).}
\label{tab:ablation_overlap}
\begin{tabular}{lcc}
\toprule
\textbf{Variant} & \textbf{HR@1000} & \textbf{Rel. Change} \\
\midrule
No Overlap & \textbf{0.3306} & - \\
\midrule
Temporal Overlap & 0.3283 & -0.7\% \\
\bottomrule
\end{tabular}
\end{table}

The performance drop is negligible (0.3306 $\to$ 0.3283). This demonstrates high robustness: the Decoder's attention mechanism automatically learns to ignore redundant information in the Memory if it is already present in the recent interactions.

\begin{figure}
    \centering
    \includegraphics[width=1.\linewidth]{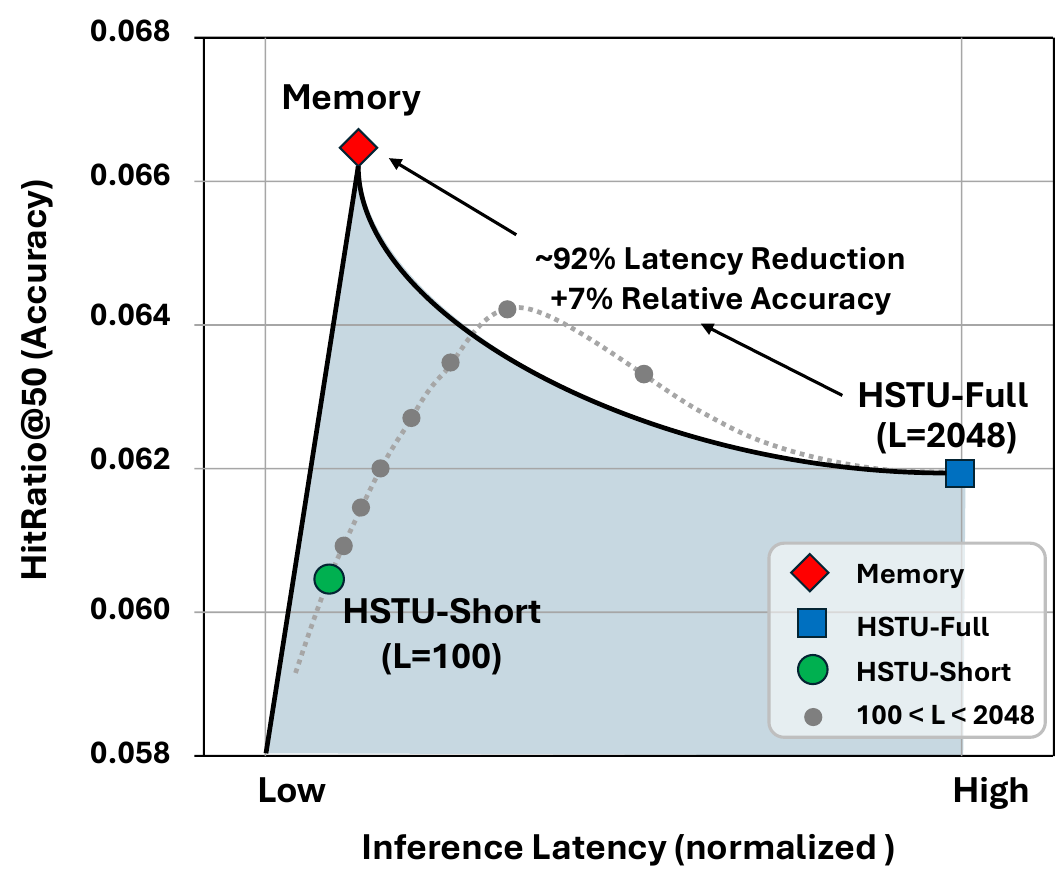}
    \caption{Efficiency-Accuracy Trade-off.}
    \label{fig:PFM_pareto}
\end{figure}

\subsection{Efficiency Analysis: The Pareto Frontier}

To visualize the trade-off, we plot Inference Latency vs. HitR@50 (Figure ~\ref{fig:PFM_pareto}). HSTU-Full occupies the high-accuracy, high-latency region, while HSTU-Short occupies the low-accuracy, low-latency region. Compressing user historical interaction sequences into Preference Memory dominates the Pareto frontier, achieving 107\% of HSTU-Full's accuracy while incurring only ~8\% of the inference latency. This drastic efficiency gain allows for the deployment of larger, deeper models within the same latency budget.

\begin{figure*}
    \centering
    \includegraphics[width=1.\linewidth]{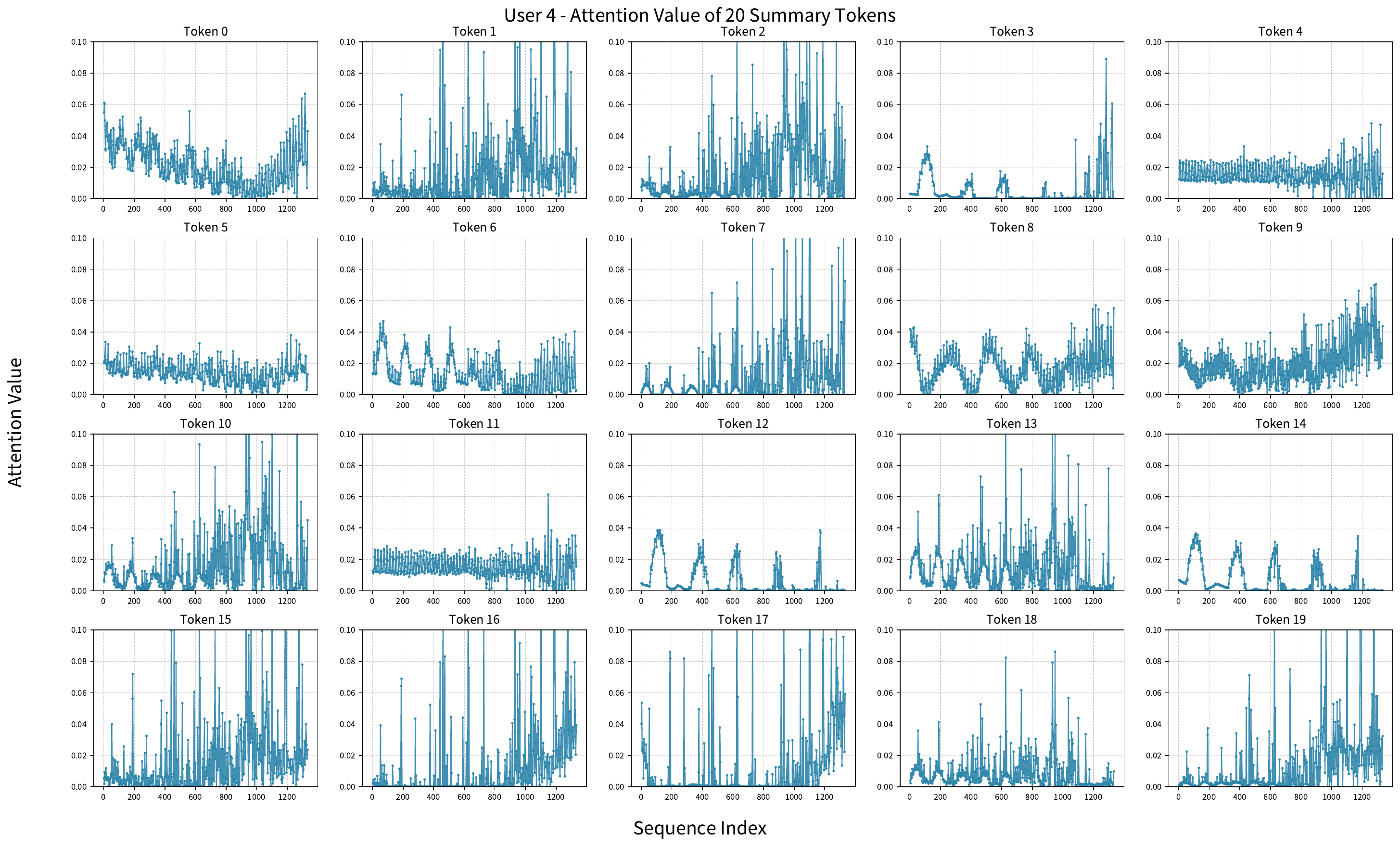}
    \caption{\textbf{Temporal Disentanglement of Preferen.} The attention weights across sequence positions reveal specialized temporal roles: \textbf{Token 0} retains early history (User Identity), \textbf{Tokens 16/19} focus on recent interactions (Working Memory), and \textbf{Tokens 3/14} capture diverse periodic patterns (Long-term Habits).}
    \label{fig:User_4_Attention_Line}
\end{figure*}

\begin{figure*}
    \centering
    \includegraphics[width=1.\linewidth]{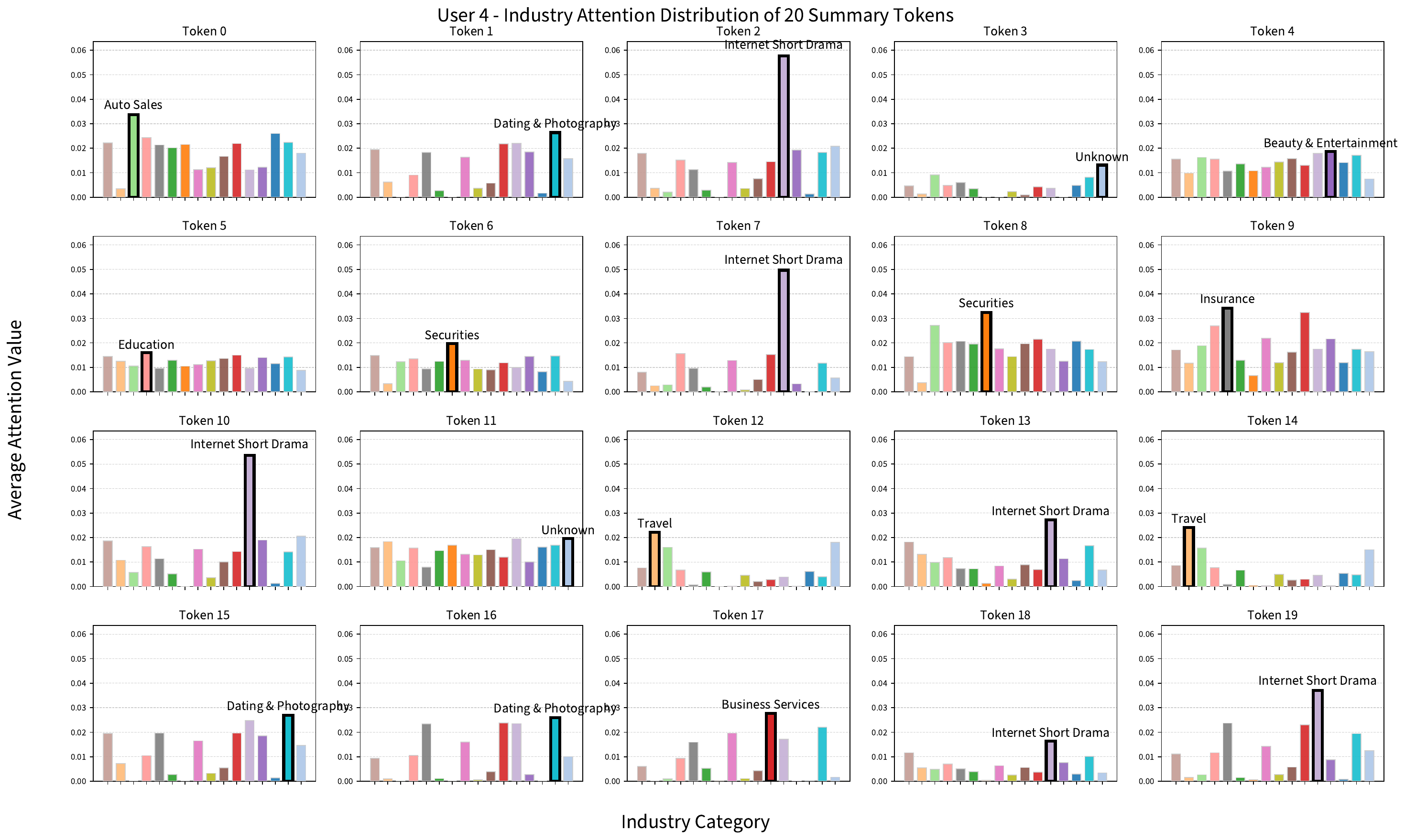}
    \caption{\textbf{Semantic Specialization and Orthogonality.} The attention distribution across item categories demonstrates that memory slots evolve into \textbf{``Domain Experts''} (e.g., Token 10 for Social Drama, Token 17 for Business). The sparsity of the attention weights confirms that Preference Memory learns a disentangled and noise-robust representation of user interests.}
    \label{fig:User_4_Industry_Attention_Distribution}
\end{figure*}
\subsection{Visualization of Preference Memory}

To interpret the internal mechanisms of Preference Memory, we visualize the attention weights of the $Q_{mem}$ across sequence positions (Figure ~\ref{fig:User_4_Attention_Line}) and item categories (Figure ~\ref{fig:User_4_Industry_Attention_Distribution}). The results demonstrate that Preference Memory automatically disentangles user history into distinct temporal and semantic patterns.

Temporal Specialization (Figure ~\ref{fig:User_4_Attention_Line}): The attention distributions reveal that $Q_{mem}$ spontaneously adopt specialized temporal roles

\begin{itemize}
    \item \textbf{Recent Attention}: Tokens 16, 17, and 19 exhibit strong attention peaks at the sequence tail. These queries capture the immediate intent shifts and the most recent user context.
    \item \textbf{Early History Retention}: In contrast, Token 0 focuses almost exclusively on the sequence start (indices 0-200). This suggests it retains ``first impressions" or foundational user attributes, preventing the forgetting of core user identity.
    \item \textbf{Periodic Patterns}: Tokens 3, 6, 8, 12, and 14 show distributed spikes across the entire history. Notably, they capture different frequencies—some sparse (Token 3) and others dense (Token 6)—indicating the model tracks diverse recurring habits rather than scanning uniformly.
    \item \textbf{Hybrid Functionality}: Token 7 demonstrates a fusion of roles. It displays both periodic historical spikes and a surge in recent attention, effectively bridging long-term patterns with immediate relevance.
\end{itemize}

Semantic Specialization(Figure ~\ref{fig:User_4_Industry_Attention_Distribution}): The memory slots also achieve high-level semantic disentanglement, evolving into ``domain experts" for specific categories

\begin{itemize}
    \item \textbf{Domain Expertise}: Specific tokens evolve into ``domain experts" with highly concentrated attention. For instance, Token 10 dedicates nearly all its attention mass to ``Internet Short Drama," while Token 8 specializes in ``Securities" and Token 9 in ``Insurance".
    \item \textbf{Sparsity and Noise Filtering}: Crucially, these tokens maintain lower weights for unrelated categories. This sparsity confirms that the limited memory capacity (C=20) forces the model to strictly filter noise, retaining only the most salient semantic signals via the Information Bottleneck principle.
    \item \textbf{Orthogonal Representation}: The distinct semantic focus of these ``experts" suggests that Preference Memory learns a disentangled basis for user interests. By minimizing redundancy between slots, the model constructs a complex user profile through the composition of independent attributes, effectively avoiding interference between diverse topics.
\end{itemize}

Conclusion: In summary, the Preference Memory go beyond simple compression to structurally organize user history. By dynamically assigning specialized roles—ranging from temporal anchors to semantic experts—the model constructs a compact, disentangled, and comprehensive representation of lifelong interests.

\end{document}